\documentclass{article}
\usepackage{glas}

\def\wma{${\rm sin^2} \theta_{\rm W}^{\rm eff}$}
\def\sm{{\sffamily SM}}
\def\ppt{$(p, \, p_{\rm T})$}
\def\afbb{$A_{\rm FB}^b$}
\def\afbc{$A_{\rm FB}^c$}
\def\st{\scriptstyle}

\def\be{\begin{equation}}
\def\ee{\end{equation}}
\def\bea{\begin{eqnarray}}
\def\eea{\end{eqnarray}}

\begin{document}

\begin{titlepage}{GLAS-PPE/1999--02}{March 1999}

\title{HEAVY QUARK ASYMMETRIES AT {\sffamily LEP}}

\author{ANDREW W. HALLEY}

\note{{\sffamily CERN}, {\sffamily EP} Division, CH-1211, Geneve 23,
Switzerland \\ (on leave of absence from the University of Glasgow,
Scotland, UK)}

\collaboration{on behalf of the {\sffamily LEP} Collaborations}

\begin{abstract}
Measurements of $b$ and $c$ quark
asymmetries using data collected at {\sffamily LEP 1} are described.
The relative merits of each of the individual techniques used is
emphasised as is the most profitable way of combining them.  Effects
of radiative corrections are discussed, together with the impact of
these measurements on global electroweak fits used to estimate the
expected mass of the Higgs boson. 
\end{abstract}
\end{titlepage}
%
\section{Motivation for Measuring Quark Asymmetries}
The accurate determination of the \mbox{forward-backward} asymmetries,
$A_{\rm FB}$, of quarks serves to test the structure of Standard Model
({\sffamily SM})\cite{SM} couplings to fermions. They also probe 
radiative corrections to the {\sffamily SM} and consequently
allow greater precision when predicting unknown parameters of the
model.  These are increasingly used to constrain
uncertainties on the mass of the Higgs boson\cite{lepewwg}.

Global fits to electroweak data assume the {\sffamily SM} structure of Z
couplings to leptons ($e, \, \mu \, \tau$) and both up ($u, \, c$) and
\mbox{down-type} ($d, \, s, \, b$) quarks.  Given recent, highly
accurate, lepton measurements from the $\tau$
polarisation\cite{taupol} and purely leptonic \mbox{forward-backward}
asymmetries\cite{lepewwg}, a similar precision in the quark sector
is needed to confirm the internal consistency of the model.

The suite of complimentary measurements described here provide such a
precision for both up and \mbox{down-type} quark families.  Performing
these measurements at {\sffamily LEP 1} offers several advantages.  The
sensitivity of initial state couplings to the effective weak
mixing angle, ${\rm sin^2}\theta_{\rm w}^{\rm eff}$, is compounded by
large, measurable asymmetries from quark final states close to the
Z.  Heavy quark asymmetries in particular are especially favourable, as
flavour and direction of the final state quark can be tagged
with greater ease than is the case with lighter quarks.
%
\section{Definitions and Experimental Issues}
In the \sm, the differential \mbox{cross-section} for the process $e^+
\, e^- \rightarrow f^+ \,f^-$ can be written as~:
\begin{equation}
\frac{1}{\sigma} \: \frac{{\rm d} \, \sigma^f}{ {\rm d} \, cos \,
\theta} \: = \: \frac{3}{8} \; ( 1 + cos^2 \theta) \; + \; 
A_{\rm FB}^f \, cos \, \theta
\end{equation}
where $A_{\rm FB}^f$ defined to be the \mbox{forward-backward}
asymmetry for fermion flavour, $f$.  It can be expressed as~:
\begin{equation}
A_{\rm FB}^f \: = \: \frac{3}{4} \, {\cal A}_e \; {\cal A}_f
\end{equation}
where ${\cal A}_f$ is the polarisation of the fermion concerned~:
\begin{equation}
 {\cal A}_f \: = \: \frac{ 2 x}{1 \, + \, x^2 } \: = \: 
1 \, - \, \frac{ 2 q }{I_3^f} \, ( {\rm sin^2} \theta_{\rm W}^{\rm
eff} \: + \: {\cal C}_f )
\end{equation}
where $x$ is  the ratio of the vector and axial couplings of the
fermion to the Z.  This final form, separates the terms containing
sensitivity to parameters of the \sm, such as $( m_{\rm t}, \, m_{\rm
H})$ through \wma, from vertex corrections in ${\cal C}_f$.  The
latter are typically of the order of $\sim 1\%$ for $b$
quarks\cite{physics:at:lep}.  

For hadronic decays of quarks, the precise direction of the final
state fermion is not accessible experimentally, and so the direction of
the thrust axis is usually signed according to methods of correlating
the charge of the quark its final decay products.  Asymmetries
are of the order of $\sim 10\%$ for $b$ and $c$ quarks but are diluted
by several effects.  These are caused primarily by the
correlation method mistagging the quark charge, or by $B^0 \bar{B^0}$
mixing or cancellations between other quark backgrounds.

Consequently, the methods of measurement described here represent
different compromises between rates of charge mistag, and the
\mbox{efficiency$+$purity} of the flavour tagging procedure.  With the
increasing sophistication of analyses, several methods of tagging
the charge and flavour of decaying quarks are available.  These are
applied either singly or in combinations.

Minor complications arise when interpretating the results of these
analyses, in the form of corrections to pure electroweak predictions
of heavy quark asymmetries.  For example, quark mass corrections to
the electroweak process are generally small, {\em eg.} representing
shifts of 0.05\% in the calculation of $A_{\rm FB}^b$, and are well
understood theoretically\cite{physics:at:lep}.  Larger and more
problematic corrections arise from hard gluon emission in the final
state.  These \mbox{so-called} {\sffamily QCD} corrections are mass, flavour
and analysis dependent and so their treatment and associated
uncertainties must be handled with care.
%
\section{Latest Techniques for Measuring \afbb}
\subsection{Semileptonic Decays of Heavy Quarks}
\label{semilepton}
The ``classical'' method of measuring heavy quark asymmetries relies
on differences in the momentum and transverse momentum, \ppt,  spectra
of leptons arising from semileptonic $b$ and $c$ quark decays.  This
method benefits from an unambiguous charge and flavour tag in the case
of unmixed $b$ and $c$ hadrons.  It however suffers from several
disadvantages when used on a more general sample.

The methods reliance on a pair of correlated inputs, such as
lepton \ppt alone, leads to a dependence of the method on the precise values
of the branching ratios, \mbox{BR($b\rightarrow l$)} and 
\mbox{BR($b\rightarrow c \rightarrow l$)}, semileptonic decay
modelling and $c$ branching fractions assumed in Monte Carlo
simulations.  Some uncertainties are constrained by full use of
measurements from lower energy experiments, according to the
prescription detailed in\cite{lepewwg}.  However, the extrapolation
of such measurements to {\sffamily LEP} energies, and into the different
environment of a fragmented jet containing a \mbox{$b$-hadron}, leaves
residual systematic uncertainties.  As the purity of the sample, and
its charge mistag, are determined by Monte Carlo simulation alone
these residual uncertainties are correlated between the 4 {\sffamily LEP}
experiments.

The relatively small value of the ($b\rightarrow l$) branching
fraction means that an overall $b$ tagging efficiency of the order of
$\sim 13\%$ is typically achieved with reasonable purities of {\em
eg.} $\sim 80\%$ in the case of\cite{aleph:leptons}.  
Effects of $B^0 \bar{B^0}$ mixing, ``cascade'' $b$ decays ($b
\rightarrow c \rightarrow l$), backgrounds from $c$ decays ($c
\rightarrow l$) and other light quark sources, including detector
misidentification, all serve to dilute the otherwise excellent mistag
rate.

The systematic impact of such effects can be severely reduced by
fitting simultaneously for the the \mbox{time-integrated} mixing
parameter, $\bar{\chi}$\footnote{$\bar{\chi}$ is defined to be the
probability that a $B$ meson has oscillated to a $\bar{B}$ meson by
the time of its decay.}, together with the $b$ and $c$ asymmetries in
a ``global'' fit.  Otherwise both $\bar{\chi}$ and \afbc \hspace{0.5mm} are  
inputs to the determination of \afbb.  Their values and uncertainties
are either fixed by experiment or, in the case of \afbc,  set to their
\sm \hspace{0.5mm} expectation and the dependence on \afbb used as
input to subsequent electroweak ({\sffamily EW}) fits\cite{lepewwg}. 
Uncertainties can be further minimised by adding input information
which discriminates between $b$ and $c$ events.  In such
analyses\cite{delphi:leptons,opal:leptons} information coming from
lifetime tags, using silicon vertex detectors ({\sffamily VDET's}), and
event shapes amplify the discrimination from semileptonic \ppt
spectra.

\begin{figure}[h]
 \begin{center} 
  \mbox{\epsfig{file=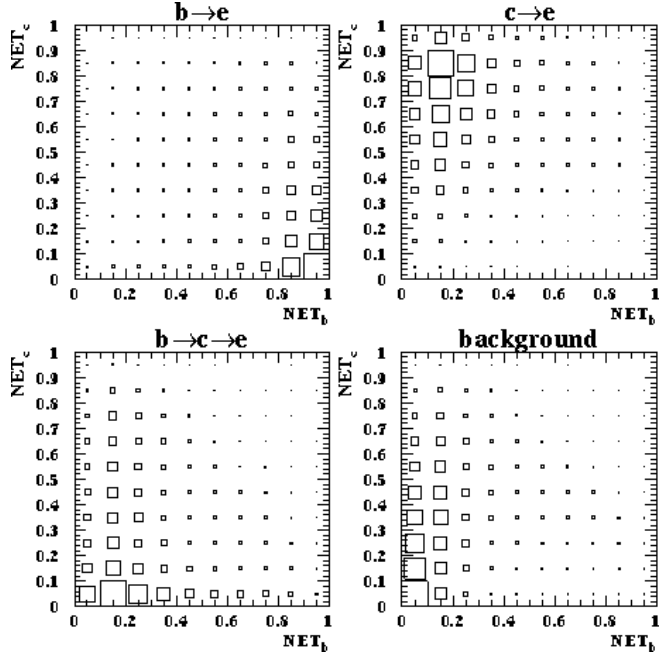,width=8.6cm}}
 \caption{\em Two-dimensional summary plot of the $c$ tag neural net
output vs. that of the $b$ tag {\sffamily NN} output for each of the
expected lepton sources in the {\sffamily OPAL} semileptonic asymmetry
analysis. The areas of displayed pixels is proportional to the bin contents.}
 \label{figure-opal nnout}
 \end{center} 
\end{figure}

The {\sffamily LEP} experiments make use of such additional inputs, and
extra fit quantities, to varying degrees as is summarised in
Table~\ref{table-lepton fits}.
\begin{table}[h]
 \begin{center}
  \begin{tabular}{|c|cccc|} \hline
{\em Inputs} & {\em lepton \ppt} & {\em lifetime} & {\em jet shapes} &
$\bar{\chi}$ \\
{\em Expt.} & {\bf ADLO} & {\bf DO} & {\bf DO} & {\bf DL} \\
\hline
{\em Outputs} & \afbb & $\bar{\chi}$ & \afbc &  \\ 
{\em Expt.} & {\bf ADLO} & {\bf AO} & {\bf DO} & \\
\hline
  \end{tabular}
  \caption{\em Additional inputs and fit parameters for the
semileptonic decay measurements of \afbb from the 4 {\sffamily LEP}
collaborations {\sffamily ALEPH, DELPHI, L3} and {\sffamily OPAL} ({\sffamily ADLO}).}
  \label{table-lepton fits}
 \end{center}
\end{table}
Analyses which make use of the classical semileptonic \ppt spectra,
and complement it with lifetime tags in this way, are the most
powerful, statistically and with greater systematic control.  For
example, the {\sffamily OPAL} measurement\cite{opal:leptons} uses lepton
\ppt and \mbox{event-shape} information as inputs to a neural net
({\sffamily NN}) $b$ tagging algorithm.  In addition, it utilises a largely
orthogonal $c$ tag, based on lifetime information of jets in the event,
combined with the impact parameter significance and detector
identification criteria of the lepton.  The output of these two
neural nets is shown in Figure~\ref{figure-opal nnout}
where the strong separation between sources of leptons in hadronic
events is clearly evident.  The separation between $b$ and $c$ lepton
sources, and the more limited distinction between those and other
background sources, enables both an precise determination of \afbb,
$\bar{\chi}$ and the most accurate measurement of \afbc from the same
sample of events.  The net gain of such a method is an approximate
$\sim 25\%$ improvement in the statistical
sensitivity of \afbb and a $\sim 25\%$ improvement in that of \afbc.
A summary of the current results for \afbb \hspace{0.5mm} from
semileptonic measurements at {\sffamily LEP} is given in
Table~\ref{table-basyms}.
\begin{table}[h]
 \begin{center}
  \begin{tabular}{|l|llll|} \hline
\multicolumn{5}{|c|}{\em Semileptonic Measurements of \afbb} \\ \hline
{\em Experiment} & \afbb & {\em stat.} & {\em syst.} & {\em total} \\
\hline
{\sffamily ALEPH } & 0.0965 & $\pm$0.0044 & $\pm$0.0026 & $\pm$0.0051 \\
{\sffamily DELPHI} & 0.0979 & $\pm$0.0065 & $\pm$0.0029 & $\pm$0.0071 \\
{\sffamily L3    } & 0.0963 & $\pm$0.0065 & $\pm$0.0035 & $\pm$0.0074 \\
{\sffamily OPAL  } & 0.0910 & $\pm$0.0044 & $\pm$0.0020 & $\pm$0.0048 \\ \hline
\multicolumn{5}{|c|}{\em Lifetime and Jetcharge Measurements of \afbb}
\\ \hline
{\em Experiment} & \afbb & {\em stat.} & {\em syst.} & {\em total} \\
\hline
{\sffamily ALEPH } & 0.1040 & $\pm$0.0040 & $\pm$0.0032 & $\pm$0.0051 \\
{\sffamily DELPHI} & 0.0979 & $\pm$0.0047 & $\pm$0.0021 & $\pm$0.0051 \\
{\sffamily L3    } & 0.0855 & $\pm$0.0118 & $\pm$0.0056 & $\pm$0.0131 \\
{\sffamily OPAL  } & 0.1004 & $\pm$0.0052 & $\pm$0.0044 & $\pm$0.0068 \\ \hline
  \end{tabular} \caption{\em Summary of latest measurements of \afbb
  from the 4 {\sffamily LEP} experiments.}  
\label{table-basyms}
  \end{center}
\end{table}
%
\subsection{Lifetime Tagging and Jetcharge Measurements}
An alternative, complementary technique to measure \afbb
\hspace{0.5mm} is based on lifetime information from silicon vertex
detectors.  This is then combined with a fully inclusive charge
correlation 
method, referred to as the ``jetcharge'' technique.  This method was
initially pioneered using samples of untagged hadronic events
containing all types of quark flavours accessible at these
energies\cite{aleph:qfb}.  As a consequence of the low semileptonic
branching ratios, such inclusive measurements are almost entirely
uncorrelated from semileptonic measurements and so can either be
combined or used as a \mbox{cross-check} for consistency of
measurements between different methods.

The jetcharge method is based upon the correlation between leading
particles in a jet, with that of the parent quark.  A hemisphere based
jetcharge estimator is formed using a summation over particle charges,
$q$, weighted by their momentum, $\vec{p}$~:
\begin{equation}
  Q_{\rm F} \: = \:
  \frac{ \sum_i^{ \vec{p_i} \cdot \vec{T} > 0 }
         \mid \vec{p_i} \cdot \vec{T} \mid^{\kappa} \, q_i }
       { \sum_i^{ \vec{p_i} \cdot \vec{T} > 0 }
         \mid \vec{p_i} \cdot \vec{T} \mid^{\kappa}        } \, ,
  \label{equation-charge summation}
\end{equation}
and analogously for $Q_{\rm B}$.  The $\kappa$ parameter is used to
optimise the measurement sensitivity.  The charge flow between
hemispheres, namely $Q_{\rm FB} = Q_{\rm F} - Q_{\rm B}$ is then used
to sign the direction of the thrust axis.  Currently all {\sffamily LEP}
collaborations use this method, and the above formalism.

The method benefits from many of the systematic studies performed in
untagged samples\cite{aleph:qfb}, especially to
understand the degree of charge correlation between hemispheres in
background events and their light quark parents.  The recent \afbb
\hspace{0.5mm} 
analysis carried out by {\sffamily DELPHI} also illustrates this
method\cite{delphi:jetcha}.
In addition to jetcharge information, {\sffamily OPAL} makes uses of a
weighted vertex charge method\cite{opal:jetcha}.  This quantity is
a weighted sum of the charges of tracks in a jet which contains a
tagged secondary vertex, {\em ie~:}
\begin{equation}
q_{\rm vtx} \: = \: \sum_{{\rm tracks}=i} \, \omega_i \, q_i
\end{equation}
where $q_i$ is the charge of each track,$i$ and $\omega_i$ is
related to the probability that the tracks comes from the secondary
vertex relative to that it came from the primary.  The latter
probabilities are determined using impact parameter, momentum and
multiplicity information.  An estimate of the accuracy of the $q_{\rm
vtx}$ charge estimator is derived from its variance.

Selecting hemispheres with $\mid q_{\rm vtx} \mid > 1.4 \, \times \,
\sigma_{\rm q} + 0.2$ largely removes neutral $B^0$ mesons,
and those with poorly measured vertex charges.  This also
leads to a severe reduction in the size of the event sample, leaving
only $\sim 13,000$ events out of a total, untagged  input of
roughly 4 million hadronic events.  Hence, the contribution of the
vertex charge measurement, when combined with the jetcharge
determination of \afbb, is relatively low.

More significant improvements in statistical precision and control of
systematic uncertainties can be obtained from the variety of new
techniques summarised in Table~\ref{table-new tech}.
\begin{table}[h]
 \begin{center}
  \begin{tabular}{|l|l|} \hline
{\em Expt.} & {\em Improvement} \\ \hline
{\bf A L}  &{\rm  Fit data as a function of tag purity.} \\
\hline
{\bf A L}  &{\rm  Increase acceptance beyond central region.} \\
\hline
{\bf ADLO}  &{\rm  Fit asymmetry as a function of angle.} \\
\hline
{\bf A}  &{\rm  Fit data using range of $\kappa$ values.} \\
\hline
{\bf ADO}  &{\rm Extract $b$ quark mistag factor from data.} \\
\hline
{\bf AD}  &{\rm Extract lighter quark mistags from data.} \\
\hline
  \end{tabular}
  \caption{\em Recent improvements to the lifetime and jetcharge
method for measuring \afbb from the 4 {\sffamily LEP}
collaborations {\sffamily ALEPH, DELPHI, L3} and {\sffamily OPAL} (ADLO).}
  \label{table-new tech}
 \end{center}
\end{table}
Experiments make use of these techniques to varying
degrees, the most significant of which being the improvement in
statistical sensitivity gained by fitting to the asymmetry as a
function of angle.  With increased statistical precision, comes the
need for improved systematic control.  The most important
of these being the extraction of the charge mistag factor for $b$
quark from data.  {\sffamily ALEPH, DELPHI} and {\sffamily OPAL} now
perform this extraction while {\sffamily ALEPH} also extracts it as a
function of the polar angle of the thrust axis.  This takes into
account particle losses close to the edge of detector acceptance.

\begin{figure}[h]
 \begin{center} 
  \mbox{\epsfig{file=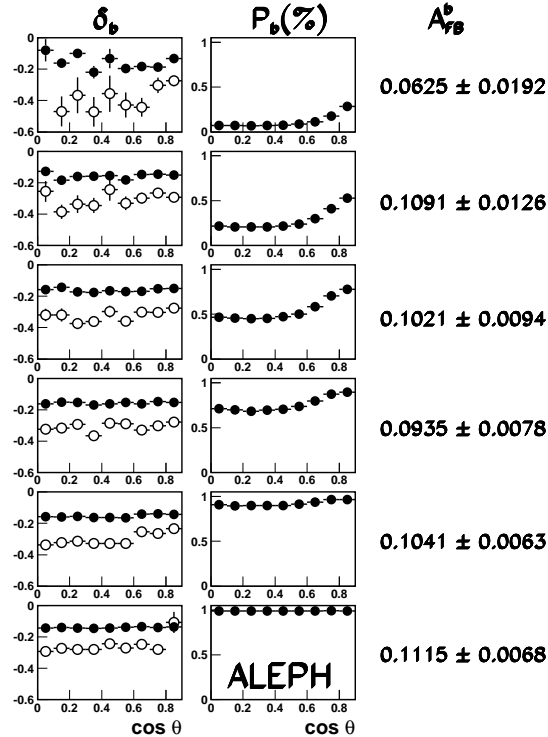,width=8.6cm}}
 \end{center} 
 \caption{\em Extracted fit variables from the {\sffamily ALEPH}
lifetime+jetcharge measurement of \afbb.  Columns represent
the $b$ charge mistag factor and sample $b$ purity as a
function of angle in samples with
increasing $b$ purity.  Filled and open circles represent $\kappa$
values of 0.5 and 2.0 in equation~(4) respectively.}  
 \label{figure-alout}
\end{figure}

The output distributions from the {\sffamily ALEPH} measurement are shown in
Figure~\ref{figure-alout}.
The asymmetry measurement is made
separately in each bin of polar angle, $\kappa$ and $b$ sample purity
before being combined.  Some points of interest in such new
analyses include the increased statistical power arising from the bins
at low angles, even those where the thrust axis lies outside the {\sffamily
VDET} acceptance.  The large value of the asymmetry in these regions
compensating for the low tagging efficiencies.  The increase in sample
$b$ purity at large $cos \, \theta$ for high $b$ purity samples, is
due to the loss of tracks at the edge of the {\sffamily VDET} acceptance.
This affects $b$ events the least, as more tracks with large $p_{\rm
T}$ to the thrust axis continue to tag the event. This is true
to a much lesser extent for lighter quark flavours.

Measurements of the $b$ asymmetry using the lifetime and jetcharge
method are also summarised in Table~\ref{table-basyms}.  It is interesting
to note that such measurements, whilst providing similar sensitivity
to \afbb as semileptonic $b$ decays,  as yet do not provide the
possibility of analogous measurements of \afbc.  
\section{Latest Techniques for Measuring the $c$ Asymmetry}
In contrast to the incremental progress in the field of $b$
asymmetries, measurements of the corresponding quantity in $c$ decays
have improved dramatically in recent years.  Of the 4 {\sffamily LEP}
collaborations, {\sffamily DELPHI, L3} and {\sffamily OPAL} have determined the
$c$ asymmetry as an output of global semileptonic fits to $b$ and $c$
decays\cite{delphi:leptons,l3:leptons,opal:leptons} whereas {\sffamily
ALEPH, DELPHI} and {\sffamily OPAL} have performed the same measurement
using fully reconstructed samples of $D$ meson 
decays\cite{aleph:D,delphi:D,opal:D}.  
As semileptonic measurements are discussed in
Section~\ref{semilepton}, only the latter are described here.

The method of exclusively reconstructing $D$ decays aim to use as many
channels as possible by reconstructing the
$D^0$ through its decay to $K^-\pi^+$.  The $D^0$ is generally
reconstructed\cite{aleph:D} by taking all 2 and 4 track combinations
and a $\pi^0$ candidate with zero total charge.  Those combinations
with odd charges are then used to form possible $D^+$ candidates.
Each experiment reconstructs a different subset out of a total of 9
different decay modes.  The dominant channels however are the $D^+
\rightarrow K^- \pi^+ \pi^+$ and $D^{*+}$ modes, with all modes offering
some statistical power.  Similarly, each experiment has
different selection criteria in each mode, depending on the momentum
and particle identification resolutions of the detectors.

\begin{figure}[h]
 \begin{center} 
\mbox{\epsfig{file=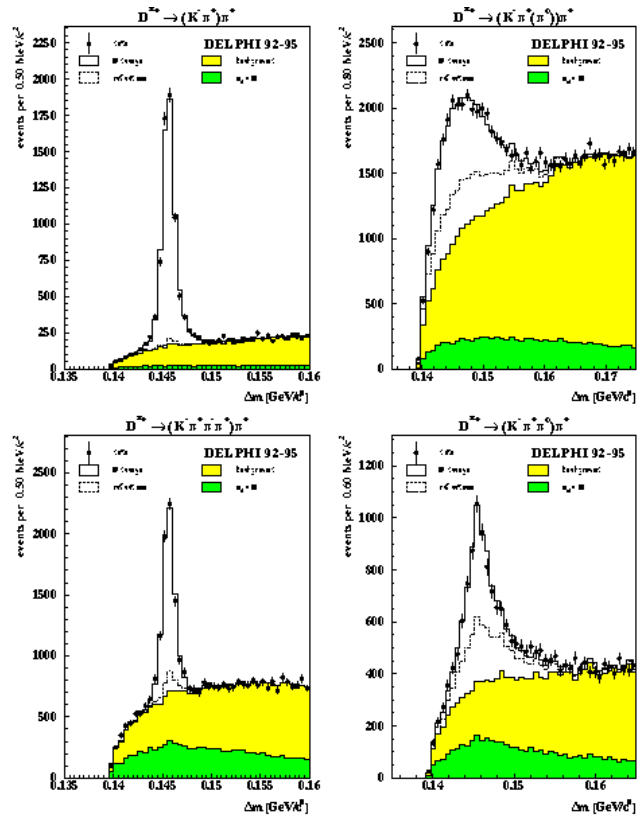,width=8.6cm}}
 \end{center} 
 \caption{\em Difference distributions between the mass of the
$D^{*+}$ and the $D^0$ candidate for different decay 
modes.}
 \label{figure-delphidmass}
\end{figure}

These differences lead to widely varying efficiencies and
\mbox{signal-to-background} ratios.  For example, the {\sffamily
DELPHI} mass difference distributions for 4 of the 8 selected modes
are shown in Figure~\ref{figure-delphidmass}.
An important advantage of such measurements is the ability to
determine the background asymmetry, $A_{\rm FB}^{\rm bkg}$, from data,
\mbox{mode-by-mode} using information from sidebands. 

A major difficulty is encountered when trying to correct for the
substantial fraction of events due to $B^0 \rightarrow D$ decays for
$B^0-\bar{B^0}$ mixing.  Each $D$ mode is corrected using an
``effective'' $\bar{\chi}$ depending on the expected fractions of
$B_{\rm d}^0$ and $B_{\rm s}^0$ decays contributing to the mode
concerned.  These effective factors are determined from\cite{pdg}
using Monte Carlo simulation and so give rise to systematic
uncertainties. 
The observed asymmetry, $A_{\rm FB}^{\rm obs}$, is then found using~:
\begin{equation}
A_{\rm FB}^{\rm obs} \: = \: f_{\rm sig} f_c A_{\rm FB}^c \, + \, 
f_{\rm sig} ( 1 - f_c) A_{\rm FB}^b \, + \, ( 1 - f_{\rm sig} ) A_{\rm
FB}^{\rm bkg} 
\end{equation}
where $f_{\rm sig}$ is the fraction of signal of
\mbox{signal+background} events and $f_c$ is the fraction of events
containing a true $D$ meson which are due to $c$ quark events.
As far as possible, the sample $c$ purities, $f_c$, are determined
from data using lifetime, mass and \mbox{event-shape} information  in
both hemispheres of events containing a $D$ tag.  

Each experiment makes use of lifetime information, with both {\sffamily
DELPHI} and {\sffamily OPAL} using the $D$ momentum and \mbox{jet-shape}
information respectively in addition, so as to disentangle the
substantial contamination from $b$ decays.  The {\sffamily DELPHI}
experiment does so in the context of a simultaneous fit  to both $b$
and $c$ asymmetries.  However, besides constraining systematic
uncertainties, the precision available of \afbb is negligible compared
to methods discussed previously.  

The $c$ asymmetry measurements described here are summarised in 
Table~\ref{table-casyms} and indicate that, despite complex systematics, such
measurements remain primarily limited by the low efficiencies and
purities of the $D$ meson reconstruction.
\begin{table}[h]
 \begin{center}
  \begin{tabular}{|l|llll|} \hline
\multicolumn{5}{|c|}{\em Semileptonic Measurements of \afbc} \\ \hline
{\em Experiment} & \afbb & {\em stat.} & {\em syst.} & {\em total} \\
\hline
{\sffamily DELPHI} & 0.0770 & $\pm$0.0113 & $\pm$0.0071 & $\pm$0.0133 \\
{\sffamily L3    } & 0.0784 & $\pm$0.0370 & $\pm$0.0250 & $\pm$0.0446 \\
{\sffamily OPAL  } & 0.0595 & $\pm$0.0059 & $\pm$0.0053 & $\pm$0.0079 \\ \hline
\multicolumn{5}{|c|}{\em Exclusive D Tag Measurements of \afbc}
\\ \hline
{\em Experiment} & \afbb & {\em stat.} & {\em syst.} & {\em total} \\
\hline
{\sffamily ALEPH } & 0.0630 & $\pm$0.0090 & $\pm$0.0030 & $\pm$0.0095 \\
{\sffamily DELPHI} & 0.0658 & $\pm$0.0093 & $\pm$0.0042 & $\pm$0.0102 \\
{\sffamily OPAL  } & 0.0630 & $\pm$0.0120 & $\pm$0.0055 & $\pm$0.0132 \\ \hline
  \end{tabular}
  \caption{\em Summary of latest measurements of \afbc
  from the 4 {\sffamily LEP} experiments.}
  \label{table-casyms}
 \end{center}
\end{table}
Systematic errors vary widely, with the \mbox{time-dependence} of the
background asymmetry remaining merely one of many dominant sources
depending on decay mode and experiment.
%
\section{Radiative Corrections to Asymmetry Measurements}
Several small corrections must be made to the $b$ and $c$ asymmetries
extracted from the described analyses.  In the case of \afbb, {\sffamily
QED} corrections for {\sffamily ISR} and {\sffamily FSR} are relatively minor,
amounting to -0.0041 and -0.00002 respectively.  Similarly,
corrections for pure $\gamma$ exchange and $\gamma-Z$ interference
diagrams give rise to a correction of +0.0003.  Such
corrections, are general in nature and so apply equally to all
analyses.

A more difficult set of corrections involves those needed to correct
for the presence of hard gluon radiation which can distort the angular
distribution of the final state quarks when compared with the pure
electroweak process.  Estimates of such corrections to heavy quark
asymmetries have been computed to first and second order in
$\alpha_{\rm s}$ both numerically\cite{lampe} and, most recently,
analytically\cite{neerven} in different scenarios for either $c$, $b$
or massless quarks.  

A common procedure for correcting and ascribing
systematic uncertainties for the {\sffamily LEP} heavy quark asymmetries has
been developed\cite{qcdcorr}.  The more recent analytical
calculations indicate several discrepancies when compared with the
numerical results.  These remain to be resolved.  Current systematic
uncertainties are determined using a procedure of comparing the
effects between first and second order in {\sffamily QCD} and by switching
between massless quarks, and assumptions for the $c$ and $b$ quark
masses.

Further difficulties arise when considering the application of such
corrections to individual analyses.  Theoretical calculations are
typically based on the direction of the outgoing quark, whereas the
analyses described here use the thrust direction.  Further, the
sensitivity to hard gluon radiation of data, containing either a
lepton of a given \ppt, a reconstructed $D$ meson or purely inclusive
events, varies dramatically.  Effects of
\mbox{non-perturbative} {\sffamily QCD} and \mbox{higher-order} effects
during hadronisation must also be evaluated.  The latter render {\sffamily
QCD} corrections both detector {\em and} analyses dependent, {\em eg.}
event shape selections implying an implicit dependence on the strength
of gluon emission.
The correction to be applied to a given analysis is derived from~:
\begin{equation}
A_{\rm FB}^{b,c} \: = \: ( 1 - {\cal C}_{b,c} {\cal S}_{b,c} ) 
\mid_{\rm no \ QCD}
\end{equation}
where ${\cal C}_{b,c}$ represents the {\sffamily QCD} correction at
parton-through-to-hadron level, and ${\cal S}_{b,c}$ is the analysis
dependent modification.  Examples of the magnitude of the {\sffamily QCD}
corrections at the theoretical and experimental levels are shown in
Table~\ref{table-qcdcorr} for the cases of \afbb, determined using 
semileptonic and lifetime+jetcharge analyses\cite{qcdcorr}.
\begin{table}[h]
 \begin{center}
  \begin{tabular}{|lc|c|c|} \hline
 & & {\em Lepton Analyses} & {\em Lifetime+Jetcharge} \\ \hline
${\cal S}_{b}$ & min. & 0.52 $\pm$0.06 &0.24 $\pm$0.46 \\
               & max. & 0.74 $\pm$0.07 &0.36 $\pm$0.32 \\ \hline
${\cal C}_{b}$ & min. & 1.54 $\pm$0.28 &0.71 $\pm$1.36 \\
               & max. & 2.19 $\pm$0.37 &1.07 $\pm$0.96 \\ \hline

  \end{tabular}
  \caption{\em Summary of {\sffamily QCD} corrections to \afbb for the
different analysis methods.}
  \label{table-qcdcorr}
 \end{center}
\end{table}
The constants are evaluated in terms of Monte Carlo simulations,
before and after experimental cuts.  The hadronisation dependence of
corrections is included as a systematic
uncertainty by comparing results from both the {\tt
HERWIG}\cite{herwig} and {\tt JETSET}\cite{jetset} models.  It is
seen from comparing parton and hadron level that the effect of
hadronisation is to reduce the magnitude of the {\sffamily QCD}
correction\footnote{It is thought that \mbox{non-perturbative} colour
reconnection effects during the shower may be responsible.}.

It is important to note that the corrections for lifetime+jetcharge
measurements are negligible.  Significant corrections are observed for
both semileptonic,and $D$ tag measurements of both \afbb and \afbc.
Jetcharge measurements are immune to such corrections as the
\mbox{$b$-quark} charge mistag factor	 is defined using
Monte Carlo with respect to the original $b\bar{b}$ quark pair
orientation, prior to gluon or final state photon radiation, parton
shower, hadronisation and $B^0 \bar{B^0}$ mixing. All these effects
are therefore included, by construction, in the analyses, as far as
they are properly modelled in the {\tt JETSET}\cite{jetset}
hadronisation model.
\section{Conclusion and Perspectives}
With the completion of {\sffamily LEP} \mbox{data-taking} at energies close
to the Z resonance in 1995, the 4 experiments ({\sffamily ALEPH, DELPHI, L3}
and {\sffamily OPAL}), have accumulated large samples of hadronic events.
From this data, the \mbox{forward-backward} asymmetry of the $b$ quark
has emerged as the most sensitive single test of the {\sffamily SM} at {\sffamily
LEP}. The complementary $c$ asymmetry measurements offer
additional precison and a new window on couplings in the
\mbox{down-type} quark family. The precision from semileptonic
measurements of \afbb is now matched by that of
\mbox{lifetime+jetcharge} measurements.  The electroweak sensitivity of 
\afbc measurements now equals that obtained from combined quark
asymmetries measured in untagged samples, highlighting the
benficial effect of flavour tagging.

However, in light of these measurements great sensitivity to the
couplings of the {\sffamily SM}, the continuing discrepancy between
electroweak results from {\sffamily LEP} and {\sffamily SLD}\cite{altarelli} make
it essential to understand whether it is due to statistical
fluctuations or systematic effects.  Separating {\sffamily LEP} measurements of
\afbb into those from the two dominant techniques, and conservatively
ignoring correlated systematic uncertainties, indicates that there is
at most a $1.2\sigma$ discrepancy between semileptonic and
\mbox{lifetime+jetcharge} measurements.  This is insufficient to
explain the {\sffamily LEP-SLD} discrepancy but indicates that care must be
taken when considering common systematics in leptonic decay modelling
and fragmentation uncertainties.
Further improvements in both $b$ and $c$ asymmetries are possible, as
both sets of measurements are still dominated by statistics. For
semileptonic analyses, the benefits of using both lifetime and lepton
information are emphasised. In the case of the lifetime and jetcharge
method, these are most likely to come in the form of improved $b$
tagging efficiencies and extensions of  tagging to lower angles.  The
situation for improvments to measurements of \afbc is more difficult
as the number of available modes is exhausted, and efficient methods
of tagging $c$ events remain to be discovered.

At this point, without the prospect of significant, further {\sffamily LEP}
\mbox{data-taking}  at the Z, it is important to focus upon the latest 
techniques which offer the greatest sensitivity to the couplings of the 
{\sffamily SM} combined with systematic control.  The measurements described
here obtain combined precisions on the $b$ and $c$ asymmetries of
2.2\% and 7.1\% respectively.  Hence, the goal of acheiving similar
precision on the Z couplings to quarks, as that obtained for leptons,
has been reached.

\end{document}